\documentclass[aps,prl,twocolumn,groupedaddress,showpacs]{revtex4}

\usepackage{graphicx}

\begin{document}

\title{Direct mechanical mixing in a nanoelectromechanical diode}

\author{Hyun S. Kim, Hua Qin, and Robert H. Blick}

\affiliation{Electrical and Computer Engineering\\ University of
Wisconsin-Madison\\1415 Engineering Drive, Madison, WI 53706, USA}


\begin{abstract}
We observe direct mechanical mixing in nanoelectromechanical
transistors fabricated in semiconductor materials operating in the
radio frequency band of $10\sim1000$ MHz. The device is made of a
mechanically flexible pillar with a length of 240 nm and a diameter
of 50 nm placed between two electrodes in an impedance matched
coplanar wave guide. We find a nonlinear {\it I-V} characteristic,
which enables radio frequency mixing of two electromagnetic signals
via the nanomechanical transistor. Potential applications for this
mixer are ultrasensitive displacement detection or signal processing
in communication electronic circuits requiring high-throughput
insulation.
\end{abstract}


\pagenumbering{arabic}

\maketitle

Mixing of electronic signals is essential for information
processing~\cite{Hagen}. A large number of electronic devices are
used for signal mixing, all of which are based on a nonlinear {\it
I-V} characteristic of one sort or another. Implementing a
mechanical mixer is advantageous because of the higher throughput
resistance, the insensitivity to electromagnetic shocks, and the
discrete resonance structure of such a mechanical system.
Conventional mechanical systems, however, have low resonance
frequencies of only some kilohertz and are thus often not applicable
for communication electronics. For these operating frequencies in
the range of some tens of megahertz to several gigahertz are
required. With the advent of nanoelectromechanical systems (NEMSs),
the resonance frequencies of mechanical systems have now been pushed
to several gigahertz simply by reducing the dimensions of the
systems~\cite{Huang,Peng}. Furthermore, we note that the integration
of field emitting devices in NEMS is very promising for mechanically
modulated millimeter wave sources. Key for functioning of these
devices is an efficient mechanism for signal mixing, as will be
demonstrated here.

 In an earlier work, we have shown that a simple
nanomechanical beam resonator driven into nonlinear response can be
used for capacitive mechanical signal mixing~\cite{Erbe_1}. In
contrast to this, we now modulate the direct current through a
nanopillar~\cite{Erbe_2} by two radio frequency signals and show
signal mixing over broad bandwidth. In Fig. 1, a typical
nanoelectromechanical single electron transistor (NEMSET) or
nanopillar is shown; the device is inserted between two electrodes
forming source and drain contacts. The pillar is defined by electron
beam lithography in silicon-on-insulator starting material, followed
by thermal evaporation of 45 nm Au, and finally applying a reactive
ion etch step to mill out the pillar (with CF$_{4}$). The oxide
thickness was 390 nm and the crystalline Si top layer thickness was
190 nm.

 For the measurements, we employ a standard dc technique to
trace at first the current versus frequency response of the system,
as shown in Fig. 2. We use an Ithaco (1210) current preamplifier in
conjunction with an EG\&G lock-in (119). The bandwidth of the
preamplifier is of the order of 20 kHz, thus ensuring amplification
of mixing frequencies below this cutoff. The frequency dependent
mixing signal can then be directly traced with the lock-in. The bias
tee on the source side of the circuit is a standard Hewlett-Packard
33150A. The direct current through the device clearly gives a
diodelike response, which depends on the mechanical response of the
system. The nanopillar itself is set into motion by applying the ac
signal leading to resonant Coulomb force (RCF) excitation at its
eigenfrequencies, as we discussed elsewhere
before~\cite{Scheible_1,Scheible_2,Scheible_3}.

\begin{figure}[!btp]
\includegraphics[width=9cm]{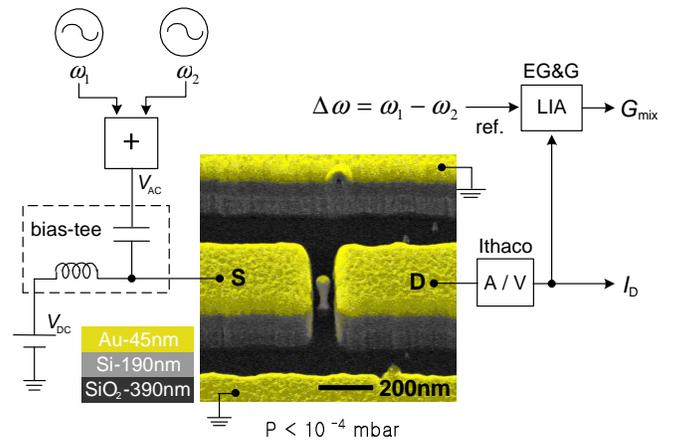}
\caption{ Schematic of the measurement circuit with a bias tee,
allowing a combination of ac and dc signals ({\it V}$_{ac}$ and {\it
V}$_{dc}$)for tracing the direct current (I$_{D}$ at drain {\it D})
frequency dependence. Two synthesizers ($\omega_{\rm 1}$ and
$\omega_{\rm 2}$) are used for direct mechanical mixing with the
difference frequency $\Delta\omega$ as a reference for the lock-in
amplifier. The nanopillar between the source and drain contacts is
dry etched from a silicon-on-insulator material and covered with a
gold top layer of 40 nm thickness (see scanning electron microscope
graph).}
\end{figure}

We find four broad resonances-with the probe station (Desert
Cryogenics TTP4) pumped to $10^{-4}$ mbar-of the nanopillar covering
the frequency range from 100 to 300 MHz (see Fig. 2). The three
different traces correspond to the three dc bias voltages from 200,
300, and finally 500 mV. The width of the resonances can be tuned by
the nanopillar dimensions. We have chosen broader resonances in
order to achieve mixing of a larger frequency range. The mechanical
and electrical responses are typically modeled by commercially
available program packages~\cite{Femlab}.

 All measurements are performed at room temperature in vacuum under (${\it P}<10^{-4}$
mbar). The inset shows the full bias dependence of the current for
on- and off-resonance positions; i.e., the ac excitation is fixed at
157, 196, and 245 MHz at a power level {\it P}=16 dBm. In other
words, the difference between the electron current on and off
resonances at ${\it V}_{\rm bias}\sim 1V$ is more than two orders of
magnitude. This is important since it indicates an excellent
signal-to-noise ratio. Even more important is the highly nonlinear
{\it I-V} characteristic, which is the necessary precondition for
signal mixing or gain when operated as a nanomechanical transistor.
\begin{figure}
\includegraphics[width=8cm]{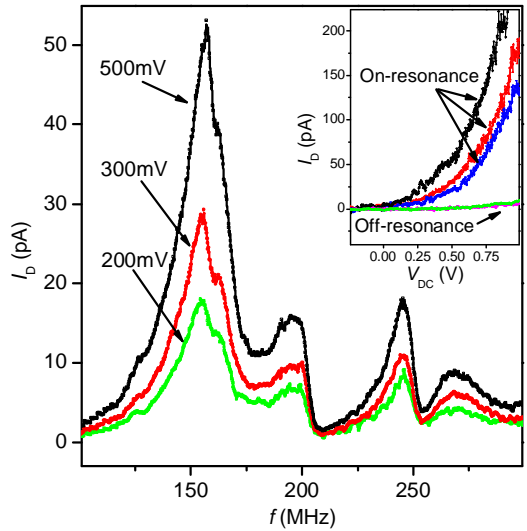}
 \caption{ Frequency spectrum of the direct current vs applied
radio frequency. All measurements are performed at room temperature
and $10^{-4}$ mbar helium gas pressure. The different traces
correspond to enhanced dc bias voltages, as indicated. Inset:
characterization of the nonlinear response of the nanopillar
compared to off-resonance current at fixed rf signal.}
\end{figure}

It has to be noted that the total current through such nanopillars
is a combination of an ordinary tunneling current and a field
emission current; it was outlined
elsewhere~\cite{Scheible_1,Scheible_2}. This effect was shown for
the lateral as well as for the vertical (nanopillar) NEMSETs.
Basically, field emission enhances conventional electron tunneling
and leads to a higher emission current. In addition, the fact that
electrons are shuttled by an island alters the classical
Fowler-Nordheim plot~\cite{Scheible_3}. This has the advantage that
field emission can be regulated by changing the island dimensions.
In addition, studying field emission through nanopillars has the
potential to reveal the intricacies of field emission since the
electron flow rate can be controlled precisely.
\begin{figure}
\includegraphics[width=8cm]{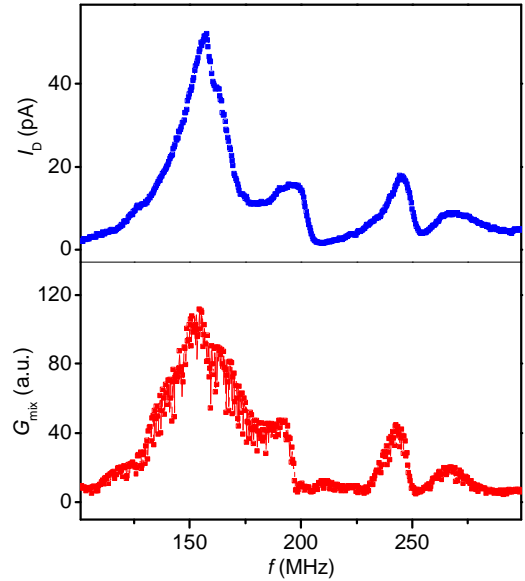}
\caption{Drain current I$_D$ vs driving radio frequency signal under
a dc bias of 500 mV (Top). Mixing signal {\it dI}/{\it dV} of the
nanopillar excited by two synthesizers running at frequencies of
$\omega_{\rm 1}$ and $\omega_{\rm 2}$/2$\pi=\omega_{\rm
1}$/2$\pi+731$ Hz (Bottom).}
\end{figure}

In order to make use of the nonlinear {\it I-V} characteristic for
signal mixing, two synthesizers~\cite{Femlab} (Agilent E8257D and HP
8656B) are applied, as shown in the circuit diagram of Fig. 1(a).
The two synthesizers are phase locked and their output signals are
combined [{\it V}$_{ac}$({\it f}1)+{\it V}$_{ac}$({\it f}2)] in an
adder and sent to the sample. The difference frequency denoted by
$\delta \omega$ =$2\pi\delta{\it f}=|{\it f}1-{\it f}2|$ is used as
a reference. This reference frequency is sent to the lock-in
amplifier and is varied from 100 Hz to 2.6 kHz, where the bandwidth
is limited by the current preamplifier and the lock-in stage. The
final current is amplified as in the standard setup; for readout of
the mixed signal, the direct current is further amplified by lock-in
amplifier operating at the reference $\delta f$. The resulting
measurements are shown in Fig. 3; here, we compare the direct
current ${\it I}_{\rm dc}$ with the mixing signal ${\it G}_{\rm
mix}$ in a frequency sweep. The mixing signal is proportional to the
second derivative of the current as
\begin{equation}\label{4}
    G_{\rm mix}\propto\hspace{0.1cm}V_{\rm ac1}V_{\rm ac2}\frac{d^2I}{dV^2}.\\
\end{equation}

As can be seen, the resonance shape is maintained, as one would
expect, which underlines the operation of the nanopillar as a
mechanical mixing element over a broad frequency range. We denoted
the mixed signal as an effective conductance {\it dI/dV} with
arbitrary units.
\begin{figure}
\hspace{1cm}
\includegraphics[width=8cm]{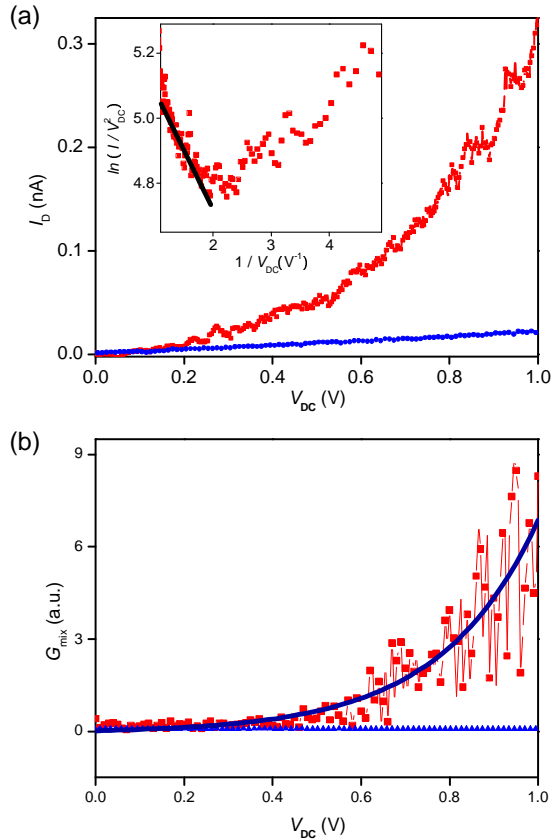}
\caption{(a) Bias dependence of the direct current through the
nanopillar and the rf driver frequencies of 157 MHz on (red) and 210
MHz off resonances (blue). (b) corresponds to the mixing signal
${\it G}_{\rm mix}$ [on (red) and off (blue)]. Inset:
Fowler-Nordheim plot with the solid line indicating a fit at high
bias. At lower bias conventional tunneling is dominant.}
\end{figure}

In Fig. 4, we compare the dc bias-dependence direct current and the
mixing signal at a specific frequency (157 MHz). In Fig. 4(a), the
current is shown on (red) and off resonances (blue); the response
for the on-resonance trace follows the modified Fowler-Nordheim
relation, we found earlier~\cite{Kim}, while that of the
off-resonance trace defines the background of tunneling electrons
from source to drain. It has to be stressed that the difference
between the on and off signals defines an excellent signal-to-noise
ratio for this single electron switch operating at radio
frequencies. Per cycle of mechanical motion of the nanopillar, it
shuttles on average $\langle{n}\rangle=\langle\it{I}\rangle/{\it
ef}$ electrons, where {\it e} is the elementary charge. For a
current of 50 pA at the resonance of 150 MHz, shown in Fig. 2, we
obtain an average number of electrons of $\sim10^3$ electrons.

 The mixing signal is presented in Fig. 4(b) for the on- and the
off-resonance cases. Again we find a stark contrast between both
traces. The inset in Fig. 4 shows the standard Fowler-Nordheim plot.
Similar to the dc results~\cite{Kim}, we find for large bias
voltages a linear dependence. The solid line indicates a fit
according to Fowler and Nordheim, while for lower bias voltages,
electron tunneling is dominant.
 In summary, we have demonstrated
mechanical mixing in a silicon nanopillar that oscillates
mechanically between two electrodes. This will have great impact for
applications such as signal processing applications and sensor
electronics. We also foresee that mechanical mixing will be
extremely important for improving measurement sensitivity of NEMS,
i.e., regarding quantum limited displacement detection, for
self-excitation of NEMS and for noise measurements on electron
shuttles~\cite{Blanter}.
\\

 This work was supported in part by the Air Force Office of Scientific Research (AFOSR)
(F49620-03-1-0420), the Wisconsin Alumni Research Foundation (WARF),
the National Science Foundation (MRSEC-IRG1), and the DARPA (NEMS
program).

\center{REFERENCES}


\end{document}